# *Coronavirus: Case for Digital Money?*

Zura Kakushadze[§¶1] and Jim Kyung-Soo Liew[†‡2]

[§] *Quantigic® Solutions LLC,[3] 680 E Main St, #543, Stamford, CT 06901*
[¶] *Free University of Tbilisi, Business School & School of Physics
240, David Agmashenebeli Alley, Tbilisi, 0159, Georgia*
[†] *SoKat Consulting, LLC,[4] Woodstock, MD 21163*
[‡] *The Johns Hopkins Carey Business School, 100 International Drive, Baltimore, MD 21202*

March 14, 2020

*"It will be like having a horse. People have horses, which is cool. There will be people who have non-autonomous cars, like people have horses. It would just be unusual to use that as a mode of transport."*
– *Elon Musk, Tesla CEO[5]*

Abstract

We discuss the pros of adopting government-issued digital currencies as well as a supranational digital iCurrency. One such pro is to get rid of paper money (and coinage), a ubiquitous medium for spreading germs, as highlighted by the recent coronavirus outbreak. We set forth three policy recommendations for adapting mobile devices as new digital wallets, regulatory oversight of sovereign digital currencies and user data protection, and a supranational digital iCurrency for facilitating international digital monetary linkages.

---

[1] Zura Kakushadze, Ph.D., is the President and CEO and a Co-Founder of Quantigic® Solutions LLC and a Full Professor in the Business School and the School of Physics at Free University of Tbilisi. Email: zura@quantigic.com
[2] Jim Kyung-Soo Liew, Ph.D., is the CEO and Founder of SoKat Consulting, LLC and an Associate Professor in Finance at the Johns Hopkins Carey Business School. Email: jim@sokat.co and kliew1@jhu.edu

[3] DISCLAIMER: This address is used by the corresponding author for no purpose other than to indicate his professional affiliation as is customary in publications. In particular, the contents of this paper are not intended as an investment, legal, tax or any other such advice, and in no way represent views of Quantigic® Solutions LLC, the website www.quantigic.com or any of their other affiliates.
[4] DISCLAIMER: This address is used by the corresponding author for no purpose other than to indicate his professional affiliation as is customary in publications. The contents of this paper are not intended as an investment, legal, tax or any other such advice, and in no way represents the views of SoKat Consulting, LLC, the website www.SoKat.co or any of their other affiliates.
[5] [Galeon, 2017].

Money is dirty.  Literally.  Surfaces of banknotes can be laden with harmful bacteria such as methicillin-resistant *Staphylococcus aureus*, *E. coli*, *Pseudonomnas aeruoginosa*, *bacillus cereus*, influenza viruses, yeast, fungi, human excreta, mold, and even cocaine and heroin.[6]  The recent coronavirus outbreak (the infectious disease COVID-19 caused by the virus SARS-CoV-2 [CDC, 2020]), that originated in Wuhan, China, has put a new emphasis on disease transmission through paper money.  People's Bank of China announced mid February 2020 its new measures, to wit, disinfecting and destroying cash to help reduce the spread of the coronavirus [Yeung, 2020].  This begs the question: *Does the latest coronavirus outbreak make yet another case for digital money?*

In this age of 5G internet and so many aspects of our lives gone digital and mobile, paper money (not to mention coins) seems archaic, if not barbaric.  At the 2017 National Governors Association Summer Meeting, Tesla CEO Elon Musk compared driving a non-autonomous car in 2037 to riding a horse [Galeon, 2017].  Unlike autonomous cars, switching to purely digital money is not associated with direct human safety concerns.  Quite the contrary, ridding the society of paper money (and coinage) would get rid of a ubiquitous medium for spreading germs.  So, what are we – the society – waiting for?

There is no technological obstacle to adopting digital money – only political and regulatory adjustments are required.  Carrying cash in a physical wallet is not any safer than "carrying" digital currency on a mobile device.  Yes, a mobile device can be lost or stolen, but so can be a physical wallet.  When the latter occurs, credit and debit cards can be misused in a fraudulent manner all the same; however, the issuers have some safeguards for that.  The loss of cash is limited to the amount in the wallet, but unlike a credit or debit card issuer, the Federal Reserve (or, more generally, a central bank) or the Treasury cannot protect the carrier of cash against such "fraudulent use" (that is, theft); nor is the carrier protected against pathogens such as the coronavirus SARS-CoV-2, possibly lurking on the banknote surfaces.  And there is no reason why risks cannot be mitigated in digital wallets by setting certain limits, akin to carrying a fixed amount of cash in one's physical wallet – with the added bonus that the carrier of digital cash never has to touch anything laden with pathogens.

---

[6] See, e.g., [Abrams and Waterman, 1972], [Angelakis et al, 2014], [Jenkins, 2001], [Maron, 2017], [Oyler, Darwin and Cone, 1996], [Pope et al, 2002], [Tanglao, 2014], [Thomas et al, 2008], [Vriesekoop et al, 2010].

Indeed, unlike exchanging cash, exchanging digital money requires no physical contact whatsoever, be it peer-to-peer, customer-to-vendor, etc. Everything can be done via scans and other touchless technologies (e.g., when paying for coffee). Even getting cash from an ATM – another archaic device – requires touching surfaces such as touch-screens and/or keypads. This too spreads germs without adding any tangible value to anyone. The World Health Organization is already encouraging using contactless payment technologies where possible as banknotes may be spreading the coronavirus [Gardner, 2020], [Huang, 2020]. So, here is our

*Policy Recommendation No. 1*: *Adapt the framework of a mobile device (such as a mobile phone) as the new digital wallet with simple and easy-to-follow rules that encourage peer-to-peer, customer-to-vendor and other digital transactions involving mobile devices.*[7]

Digital money with touchless mobile payments is the future, just as autonomous vehicles. ATMs and physical cash are the past, just as horses. However, unlike Elon Musk's aforesaid autonomous vehicle vs. horse analogy, in the future there will be no physical cash. All money will be digital as from a practical viewpoint there is no reason for a hybrid system. Consider this. Unless one has nefarious motives such as money laundering, tax evasion, illicit drug trade and other illegal activities [Rogoff, 2016], there is no compelling reason to cling to cash. Yes, government-issued purely digital money (see below) will give more control to the government. However, freedom cannot be unlimited – we do not live in an anarchy. Certain freedoms must be sacrificed for the greater good. And if ridding the society of physical cash can save lives, depriving all kinds of shady characters of their ability to use cash for illegal purposes is a small price to pay.[8] Does anyone really believe that, say, 100 years from now we will still have paper money? And, per Elon Musk, if autonomous vehicles are the norm in 20 years from now, then it is difficult to believe that purely digital money will not be.

One may argue that cash – especially in smaller denominations – is still useful for making anonymous (e.g., "vice-related", but otherwise innocuous) purchases (see, e.g., [Solomon, 2017]). Indeed, the populace's liberal freedoms should not suffer unnecessarily. However, one can imagine various simple solutions, such as small-denomination digital gift cards without the bearer's identity. Such an anonymous gift card will have to be purchased through a regular digital wallet and can be argued to still be traceable. However, so are cash withdrawals from ATMs. Sweden will be cashless by 2023 [Finextra, 2019], but not vice-free.

---

[7] For a discussion of Fintech/mobile payment adoption across different regions, see, e.g., [IMF and WBG, 2019].
[8] Prof. Kenneth Rogoff has argued that a cashless society will also benefit the poor (see, e.g., [Solomon, 2017]).

To be clear, by digital money we do not mean Bitcoin [Nakamoto, 2008] and other decentralized cryptocurrencies, or Libra-like [The Libra Association, 2019][9] cryptocurrencies issued by a consortium of private (including for-profit) entities. Instead, we mean digital money issued by sovereign governments (see, e.g., [Kakushadze and Liew, 2018] and references therein), and a universal digital iCurrency proposed in [Kakushadze and Liew, 2015, 2018] (also see [Kakushadze and Yu, 2019]) issued by a broad consortium of sovereign governments. Thus, realistically, mass-adoption of digital money is difficult to imagine without it being issued by a government (or a governmental agency) and having an appropriate regulatory oversight (in the case of national digital currencies such as the rumored-soon-to-be-forthcoming "digital Yuan" of China),[10] or by a broad consortium of governments for supranational adoption [Kakushadze and Liew, 2018], [Kakushadze and Yu, 2019]. One glaring advantage of government-issued vs. decentralized digital currencies is no need for energy-wasteful mining[11] (as in, e.g., Bitcoin).[12]

The necessity for a regulatory oversight appears to be a foregone conclusion (see, e.g., [USHCFS, 2019]); however, some basic protections must be put in place, which brings us to our

*Policy Recommendation No. 2: A sovereign digital currency must be regulated by governmental agencies with all transactions recorded and monitored to identify nefarious actors and prevent fraudulent activities. However, the golden rule is that this data cannot be sold (including to commercial entities), or used for political gain.[13]*

Digital money would therefore seem a win-win, but there are forces out there resistant to change.[14] So, other than nefarious actors, who stands to lose most from doing away with the status quo and adopting digital currency? Table 2 in [Kakushadze and Liew, 2018] summarizes pros and cons to some stakeholders from government-issued digital currencies. Here let us mention two of the stakeholders that would be adversely affected by digital money.

---

[9] For a detailed discussion of various issues with Libra, see, e.g., [Kakushadze and Yu, 2019] and references therein.
[10] See, e.g., [Horsley, 2020].
[11] For a consortium, 100 or so members would suffice [Kakushadze and Liew, 2018], [Kakushadze and Yu, 2019].
[12] For some literature on inefficiency and unsustainability issues, see, e.g., [Bariviera, 2017], [Kostakis and Giotitsas, 2014], [Nadarajah and Chu, 2017], [Urquhart, 2016], [Vranken, 2017].
[13] The Federal Reserve is designed to be free of political influence [The Fed, 2020]. Also, see, e.g., [NTIA, 2020].
[14] Including due to the lack of knowledge of how digital money would work and the fear stemming therefrom.

First, we have commercial banks, whose existing business model is jeopardized by digital money, most notably due to lower fees.  Second, we have central banks and governments, which in many instances perceive digital currency as a threat to the existing monetary system (including through its cannibalization).  However, there is a flipside to this too: the issuer-governments stand to gain, not lose, control as – unlike with paper money – with digital money *all* transactions are recorded and the information is readily available to the authorities; albeit this information should only be used to prevent nefarious activities.[15]  In this regard, the U.S. is the sovereign state with most to lose in the process of the inevitably forthcoming avalanche of government-issued digital currencies, with a clear policy implication: adapt to the changing reality, ***issue the digital dollar sooner rather than later***, or risk being marginalized [Kakushadze and Liew, 2018] – including by China (see, e.g., [Horsley, 2020]) and other players.[16]

It is difficult to imagine mass adoption of digital currencies unless they are issued by governments (or governmental agencies) with streamlined rules and regulations.  What would be truly remarkable in this regard is a supranational digital currency, which was termed *iCurrency* in [Kakushadze and Liew, 2015], backed by a broad consortium of sovereign governments [Kakushadze and Liew, 2018], [Kakushadze and Yu, 2019].  iCurrency would allow more precise estimation of global inflation, global output, global productivity, global labor gains, and other global macroeconomic indicators, as well as GDP (gross domestic product) for various countries and regions and globally.  (GDP measured using iCurrency was termed *iGDP* in [Kakushadze and Liew, 2015].)  With low sovereign government intervention (which requires a democratic protocol for the issuing consortium of sovereign governments) and low fees for global transactions, iCurrency should be a boon to our global economy.  Perhaps somewhat ironically, government-issued digital currencies may be a step forward (not backward) in this regard, depending on how the world leaders approach this issue.  This brings us to our

---

[15] In the U.S. the Bank Secrecy Act of 1970 (or BSA) [Meltzer, 1991], among other things, requires financial institutions to file a Currency Transaction Report (CTR) with the Financial Crimes Enforcement Network (FinCEN) for all transactions exceeding $10,000.  The U.S. Government records and stores information about all such transactions internally.  This information is invaluable in light of the U.S. dollar being the dominant reserve currency [Kakushadze and Liew, 2018].

[16] The Federal Reserve has been looking into developing a digital currency in the U.S. for some time – see, e.g., [Heeb, 2019].  For the digital dollar to be competitive, reducing tax rates and not taxing non-US-derived income might be required [Kakushadze and Liew, 2018].

*Policy Recommendation No. 3: Digital monetary linkages across countries would be facilitated by a supranational digital iCurrency issued by a consortium of sovereign governments. The iCurrency transaction data would be accessible to the appropriate governmental agencies of the consortium member countries subject to the limitations set forth in our Policy Recommendation No. 2 above.*[17]

Quoting from [Kakushadze and Liew, 2015]: "If mankind is destined to make it to Mars (and beyond) and establish extraterrestrial colonies in our solar system, a universal currency devoid of government control and other forms of manipulation would appear to be a necessity, rather than wishful thinking." So, ridding our monetary system of paper money indeed appears to be a win-win: among many other things, not only would it be good for our global economy, but it would also get rid of a ubiquitous medium for spreading germs and help save lives, so we would have one fewer worry in times of crises such as the latest coronavirus outbreak.

Finally, let us address the elephant in the room as it relates to iCurrency, which is the U.S. – the hegemon in the current International Monetary System. Why would/should the U.S. want iCurrency? Actually, there are two somewhat separate issues here: i) the U.S.'s leverage in dealing with its adversaries; and ii) the U.S. dollar being the dominant reserve currency.

Thus, SWIFT (Society for Worldwide Interbank Financial Telecommunication, S.W.I.F.T. SCRL), the leading worldwide network for cross-border payments, is a cooperative society operating under the Belgian law and is owned by its member financial institutions. However, the U.S. is still able to get its way by, e.g., levying sanctions on SWIFT if the latter does not cooperate (see, e.g., [Gladstone, 2012]). In this regard, if there is a supranational iCurrency and SWIFT is replaced by another network for cross-border payments in iCurrency, the U.S.'s leverage will not be diminished as it can levy sanctions all the same.[18] The U.S.'s leverage does not lie in it directly controlling SWIFT (which it does not), or that the U.S. dollar is the dominant reserve currency. Instead, it stems from the U.S. being the largest and most stable economy, and most wish to avoid their U.S. assets frozen, travel restricted, or being otherwise sanctioned.

---

[17] As mentioned above, what we are proposing here is not adopting decentralized or Libra-like cryptocurrencies as national digital currencies or a supranational iCurrency, but government-issued and appropriately regulated sovereign digital currencies and a supranational digital iCurrency issued by a broad consortium of governments. However, regulatory sentiments around the world for cryptocurrencies are still instructive; see, e.g., [GLRC, 2018].

[18] E.g., if, hypothetically, Ripple started processing payments for Iran, it would invariably draw the U.S.'s ire.

In fact, if the transaction data for a supranational iCurrency is accessible to the member states (as set forth in our Policy Recommendation No. 3 above), the U.S. stands to gain, not lose, as it will have an (essentially) unfettered access to *all* iCurrency transaction data, globally.

Now, what about iCurrency replacing the U.S. dollar as the dominant reserve currency? After all, the U.S. enjoys the "exorbitant privilege",[19] which, for our purposes here (and following [Gourinchas and Rey, 2007]), refers to the excess return of the U.S. external assets on the U.S. external liabilities. Thus, according to [Gourinchas, Rey and Govillot, 2017], this excess return was about 2.6%/yr during 1952, Q1 to 1972, Q4, and 2.4%/yr from 1973, Q1 to 2016, Q1. However, this exorbitant privilege also has a flipside: the "exorbitant duty".[20] In a nutshell, the U.S. provides *insurance* to the rest of the world (ROW) in exchange for having the (exorbitant) privilege of paying low interest rates on its safe dollar-denominated assets. So, the exorbitant privilege is not a free lunch but an insurance (risk) premium. In times of global crisis, the U.S. dollar appreciates (owing to a flight to safety), which results in a massive wealth transfer from the U.S. to the ROW.[21] Superfluously this may appear innocuous as on average the exorbitant privilege translates into a positive return. However, the U.S. debt increases in the process as the valuations of the U.S. Treasuries do not collapse [Gourinchas, Rey and Govillot, 2017].

Another aspect is the volatility. The portfolio is not well-diversified: it is long the U.S. and short the ROW (or the other way around, depending on the trade).[22] Unsurprisingly, in times of crisis the volatility goes through the roof.[23] The ballooning U.S. debt does not help.

So, it is natural to wonder, *is the exorbitant privilege really worth it?* Does the U.S. really want to bail out the ROW in times of crisis? The status of the U.S. dollar as the dominant reserve currency also harms the competitiveness of the U.S. exports as its exchange rate is an estimated 5-10% higher than it would otherwise be [Dobbs et al, 2009]. Would it not make more sense to spread the burden during the global crises more evenly across the world as opposed to shouldering it alone while the ROW gets bailed out – in fact, gets paid – for (in many cases) causing such crises in the first instance?  The latest coronavirus outbreak and the turmoil that it has caused the financial markets was certainly not caused by the U.S.  In this regard, iCurrency would be beneficial, not harmful, to the U.S.  The stability of iCurrency[24] would have to be maintained by the entire consortium, not by the U.S. alone.  This latest crisis caused by the coronavirus outbreak might be the high time for the U.S. to reevaluate its view on the U.S. dollar as the dominant reserve currency, the exorbitant privilege and exorbitant duty and shift toward a less volatile and more sustainable (at least in terms of its ever-growing debt) regime where the ROW, many parts of which got extremely wealthy through wealth transfers from the U.S., will take its fair share of responsibility for the stability of our global economy.  And the U.S.'s hegemony does not hinge on the U.S. dollar being the dominant reserve currency, but on its stable economy, legal system, science and technology, innovation, entrepreneurship, free speech, democracy, and many other things that make the U.S. what it is.

---

[19] For a nontechnical overview, see, e.g., [Bernanke, 2016].
[20] See, e.g., [Gourinchas and Rey, 2007], [Gourinchas, Rey and Govillot, 2017], [Stavrakeva and Tang, 2018] and references therein.
[21] According to [Gourinchas, Rey and Govillot, 2017], such a wealth transfer between 2007, Q4 and 2009, Q1 amounted to about 19% of the U.S. GDP, or about 3% of GPD per quarter.
[22] This is similar to the dollar carry trade (see, e.g., [Lustig, Roussanov and Verdelhan, 2014]), which is long (short) the U.S. dollar (USD) and short (long) a diversified basket of foreign currencies (typically, with equal weights).
[23] On 3/12/2020 the VIX closed at 75.47. On 11/20/2008 the VIX closed at 80.86 (while the intraday high on 10/24/2008 was 89.53). (Source: https://finance.yahoo.com.)

To summarize, a clear and transparent ***digital monetary system*** with a robust regulatory oversight to preclude bad actors from abusing it and user data protection in place would be a boon to our global economy.  Our policy recommendations above aim at achieving this goal.

---

[24] E.g., via a band similar to that maintained by the Hong Kong Monetary Authority (HKMA) for the USD/HKD (Hong Kong dollar) exchange rate (ranging between 7.75 and 7.85, the target zone fixed by the HKMA) via essentially algorithmic monetary interventions at the boundaries [Kakushadze and Liew, 2018], [Kakushadze and Yu, 2019].